\documentclass[conference]{IEEEtran}
\IEEEoverridecommandlockouts

\usepackage{cite}
\usepackage{amsmath,amssymb,amsfonts}
\usepackage{algorithmic}
\usepackage{graphicx}
\usepackage{textcomp}
\usepackage{xcolor}
\usepackage{graphicx}  
\usepackage{multirow}
\usepackage{url}  
\usepackage{hyperref}  
\def\BibTeX{{\rm B\kern-.05em{\sc i\kern-.025em b}\kern-.08em
    T\kern-.1667em\lower.7ex\hbox{E}\kern-.125emX}}
\begin{document}

\title{Channel State Information Analysis for Jamming Attack Detection in Static and Dynamic UAV Networks – An Experimental Study\\

\thanks{This research work was partially funded by the German Federal Office for Information Security under grant 01MO23014C and the Federal Ministry of Education and Research under grant 16ES1131.}
}
\author{
\IEEEauthorblockN{Pavlo Mykytyn\textsuperscript{1}, Ronald Chitauro\textsuperscript{1}, Zoya Dyka\textsuperscript{1}{2}, and Peter Langendoerfer\textsuperscript{1}{2}
\IEEEauthorblockA{\textsuperscript{1}\textit{IHP GmbH - Leibniz Institute for High Performance Microelectronics}\\
Frankfurt Oder, Germany \\
mykytyn@ihp-microelectronics.com}}

\author{
    \IEEEauthorblockN{
        Pavlo Mykytyn\textsuperscript{1}, 
        Ronald Chitauro\textsuperscript{1}, 
        Zoya Dyka\textsuperscript{1,2}, 
        and Peter Langendoerfer\textsuperscript{1,2}
    }
    \IEEEauthorblockA{
        \textsuperscript{1}\textit{IHP GmbH - Leibniz Institute for High Performance Microelectronics, Frankfurt Oder, Germany}}
        \textsuperscript{2}\textit{BTU Cottbus-Senftenberg, Cottbus, Germany} \\
        mykytyn@ihp-microelectronics.com}
}

\maketitle

\begin{abstract}
Networks built on the IEEE 802.11 standard have experienced rapid growth in the last decade. Their field of application is vast, including smart home applications, Internet of Things (IoT), and short-range high throughput static and dynamic inter-vehicular communication networks. Within such networks, Channel State Information (CSI) provides a detailed view of the state of the communication channel and represents the combined effects of multipath propagation, scattering, phase shift, fading, and power decay. In this work, we investigate the problem of jamming attack detection in static and dynamic vehicular networks. We utilize ESP32-S3 modules to set up a communication network between an Unmanned Aerial Vehicle (UAV) and a Ground Control Station (GCS),  to experimentally test the combined effects of a constant jammer on recorded CSI parameters, and the feasibility of jamming detection through CSI analysis in static and dynamic communication scenarios.
\end{abstract}


\section{Introduction}
The rapid expansion of IEEE 802.11 networks over the past decade has revolutionized wireless communications, particularly in such applications as smart homes~\cite{yang2022_first}, Internet of Things (IoT)~\cite{Yang18_second}, industrial automation, and short-range high-throughput vehicular networks~\cite{BAUTISTA2020_third}. This can be contributed to their high throughput capabilities, ease of deployment, and increasingly growing demand for internet connectivity. However, the widespread usage and extensive deployment of these networks make them an attractive target for malicious actors, and thus, more exposed and susceptible to jamming attacks. Operational disruptions due to jamming in such applications, even for a short period, can result in significant consequences and financial losses. Hence, a timely detection and mitigation of jamming attacks is immensely important. However, resource limitations and insufficient computational power of the IoT devices comprising such networks, restrict the implementation and deployment of resource demanding machine learning (ML) mechanisms. Therefore, as an alternative, lightweight and efficient approaches to jamming detection are much needed, and require more research attention. Advancements in Multiple-Input Multiple-Output (MIMO) technology, in conjunction with Orthogonal Frequency-Division Multiplexing (OFDM), in 802.11 networks have enabled the use of Channel State Information (CSI)~\cite{article_fourth}. CSI captures detailed properties of a communication link, thereby offering a comprehensive representation of the received communication signal quality. These properties represent the combined effects of the multipath, scattering, phase shift, and signal fading with changing distance and presence of obstacles or interference in the path of the signal. While CSI has been widely used for a variety of sensing applications including human activity recognition~\cite{ZHANG2023_fifth}, gesture recognition \cite{9155539_sixth}, vital sign monitoring~\cite{10621199_seventh}, human pose estimation~\cite{10086600_eighths}, and indoor localization~\cite{Wu2018_nineth}, its potential for improving security, specifically early jamming detection, remains unexplored. This paper addresses this research gap by conducting experimental testing of the feasibility of jamming attack detection through direct analysis of the collected CSI parameters. 

IEEE 802.11 Wi-Fi networks are also employed for short-range high-throughput vehicular communication, e.g. between Unmanned Aerial Vehicles (UAVs) or from a UAV to a Ground Control Station (GCS). To experimentally test the feasibility of jamming attack detection in vehicular networks using CSI analysis, we set up a UAV-to-GCS communication link, implemented with ESP32-S3 modules from Espressif Systems. The set up communication link was utilized to record the CSI metrics under normal conditions, and under the influence of a Software Defined Radio (SDR) jammer in static and dynamic communication scenarios. Experimental testing showed that detection of a constant jammer in static communication scenarios was straightforward and feasible, while dynamic communication scenarios proved to be more complex.

The rest of this paper is organized as follows: Section II provides background on jamming attacks and related work. Section III describes the setup used for experimental testing. Section IV presents the results of the conducted experiments, and finally Section V concludes this work.

\section{Background and Related Work}

\subsection{Background on Jamming Attacks}\label{II-A}
Wireless communication networks are always at a high risk of jamming attacks, due to the propagation of the transmitted signals in an open and shared environment. These signals are exposed to external interference transmitted in both, licensed and unlicensed frequency bands~\cite{enwiki_10}. Interfering signals, intentionally transmitted to degrade or disrupt a legitimate communication, are considered as a jamming attack. Jamming attacks have been and continue to be a major topic in research on security of the wireless communication networks, due to the advancements and availability of low-cost and easy to use software defined radios. Recent developments in low-cost SDR technologies have made it fairly simple to launch jamming attacks on wireless networks without much prior knowledge. With off-the-shelf devices such as BladeRF~\cite{bladerf_11}, USRP~\cite{usrp_12}, or HackRF~\cite{hackrf_13}, ranging in price from a few hundred to a few thousand euros, the entrance hurdle to initiate a jamming attack has decreased significantly~\cite{unknown_14}. These devices are capable, adaptable, and cover a broad spectrum of radio frequencies. For instance, the USRP B210, a fully integrated single-board SDR platform, provides coverage for frequencies between 70 MHz and 6 GHz~\cite{brand_usrp_15}. A single SDR device also can be used to implement multiple types of jamming attacks with minimal efforts, requiring only software modifications through tools like GNU Radio~\cite{gnuradio_16}. 

There are several types of jammers and jamming attacks. In the physical layer, they can be split into the partial-band and broadband jammers~\cite{article_17}. Otherwise, they are usually classified by their ability to affect the wireless medium, and can be divided into proactive, reactive, constant, deceptive, random, and periodic jammers. In this work, we only focus on a constant jammer and its effects. A constant jammer is set up and programmed to continuously inject Additive White Gaussian Noise (AWGN) into the wireless channel, which affects the communication indefinitely. It can be aimed at a fraction of a communication channel or its entirety.

\subsection{Background on CSI}\label{II-B}
In IEEE 802.11 networks, CSI reflects the detailed properties of a communication channel and characterizes its path in a physical environment where reflections, deflections, interference and scattering may occur. These properties include time delay, phase shift, and amplitude attenuation with each change of the signal path due to obstacles or interference along the path of the radio waves. CSI takes advantage of the OFDM, employed by the IEEE 802.11 standard. OFDM is a digital multi-carrier modulation technique that splits the available frequency spectrum of the communication channel into multiple orthogonal subcarriers. Each subcarrier transports a portion of the packet with its payload, thus allowing for efficient data transmission. The amplitude of each subcarrier can be seen as the strength of the signal on that particular frequency. Analyzing the amplitude development of the subcarriers provides insights into how each frequency component of the channel is behaving. Through employing OFDM and MIMO, the IEEE 802.11 protocol can dynamically adjust transmission parameters such as modulation and coding, to improve throughput, reliability, and overall transmission performance. Transmitted signals often take several routes to reach the receiver, due to reflections, diffractions, and scattering, which is known as a multipath effect. Each route is affected in its own way by the environment and represents a different amount of time delay, amplitude decay, and phase shift. To differentiate each path, the wireless channel is characterized using a multipath model, where each path is represented by an individual component in the Channel Impulse Response (CIR), defined in~\eqref{eq:CIR}, where: 

\begin{description}
    \item[$N$]- is the total number of multipath components.
    \item[$\alpha_i$]- is the amplitude attenuation of the component $i$.
    \item[$e^{j}$]- is the complex exponential form representation.
    \item[$\theta_i$]- is the phase shift introduced by the path $i$.
    \item[$\tau_i$]- is the time delay associated with the path $i$.
    \item[$\delta(\tau)$]- is the Dirac delta function~\cite{Wu2018_nineth}.
\end{description}

\begin{equation}
    h(\tau) = \sum_{i=1}^{N} |\alpha_i| e^{-j\theta_i} \delta(\tau - \tau_i)
    \label{eq:CIR}
\end{equation}

By applying the Fast Fourier transformation (FFT) to the CIR, we can obtain the Channel Frequency Response (CFR) in the frequency domain as shown in~\eqref{eq:fft}.

\begin{equation}
    H(f) = \text{FFT} \left[ h(\tau) \right]
    \label{eq:fft}
\end{equation}

The CFR, derived using FFT from the CIR, is then used to extract subcarrier-specific amplitude and phase information. While the derivation of CIR and CFR is independent of the modulation scheme, on commercial devices it is commonly implemented together with specific modulation schemes. For the devices employing OFDM, the amplitude attenuation and phase shift can be sampled from CFR at predetermined intervals at subcarrier level, as shown in~\eqref{eq:subcarrier_response}, where:

\begin{description}
    \item[$|H_i|$]- is the amplitude attenuation at subcarrier $i$.
    \item[$\angle H_i$]- is the phase shift at subcarrier $i$.
    \item[$e^{j}$]- is the complex exponential form representation.
\end{description}

\begin{equation}
    H_i = |H_i| e^{j\angle H_i}
    \label{eq:subcarrier_response}
\end{equation}

Recent advances in wireless transmission technology have made it possible to obtain a sampled version of CFR on commercial-off-the-shelf devices, such as ESP32-S3. The CFR values, sampled at different subcarriers, reflect the amplitude and phase of the signal, presented as a series of complex numbers, which are collectively referred to as channel state information. CSI logs amplitude fluctuation, phase shift, and time delay variation (jitter), for each subcarrier in the communication channel, and each successfully received packet, providing a more detailed view of the signal quality compared to traditional metrics like RSSI and SNR~\cite{ghany_18}. Hence, partial interference and jamming can be observed through the analysis of the changes in amplitude values of individual subcarriers, increased time delay for each received packet, and the occurrence of distorted or lost packets.

\subsection{Related Work}\label{II-c}

Whilst the topic of jamming detection is not new and has been previously studied in the literature, to the best of our knowledge, there are no previous works on jamming detection in IEEE 802.11 networks using CSI analysis. Henceforth, our work is aiming to addresses this research gap and explore alternative approaches to the problem of jamming detection, which do not require high computational resources or installation of additional hardware. Nonetheless, some works that focused on other approaches of jamming detection, can also be considered relevant for our work. The authors in~\cite{article_19}, for instance, presented a real-time data-driven jamming detection approach for static IoT networks, able to identify attacks on multiple channels in 2.4 GHz band simultaneously. They sampled the values of throughput, Packet Delivery Ratio (PDR), delay, and RSSI in the regular state to generate threshold values which were used for outlier and anomaly detection. The regular state threshold values were then compared with the values during the jamming experiments in real-time. However, their work focused only on static IoT networks and not vehicular networks. The authors in~\cite{article_20} suggested a Long Short-Term Memory (LSTM)-based model for jamming detection in IoT that uses detection criteria from transport and application layers. It is based on the experimental finding that jamming attacks on the physical layer trigger cascading failures at the upper layers of the protocol stack, such as transport and application layers. However, similarly to the previous work, their detection mechanism is designed for static client-server IoT device pairs.

\section{Experimental Setup and Configuration}

\subsection{UAV-to-GCS Communication}\label{III-A}
\begin{figure}[htbp]
\centerline{\includegraphics[scale=0.175]{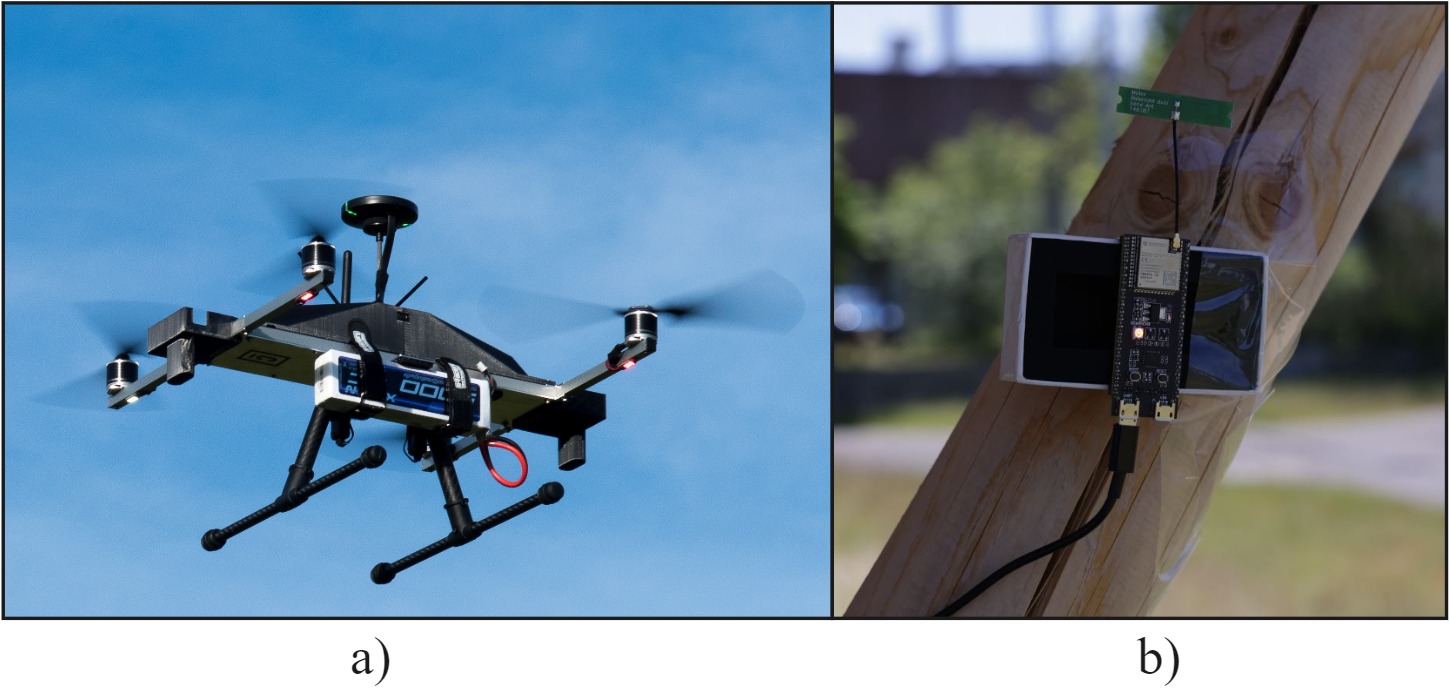}}
\caption{UAV (a) to GCS (b) communication link, set up for experimental testing.}
\label{fig:drone_base_fig}
\end{figure}

The communication between a UAV and a GCS was realized using a pair of ESP32-S3 System-on-a-Chip (SoC) with dual-core 32-bit Xtensa LX7 MCU operating at 240 MHz from Espressif Systems, with integrated Wi-Fi and Bluetooth Low Energy (BLE) connectivity. They were set up as a transmitter and receiver pair, to communicate and capture the CSI parameters from each successfully delivered packet. The first module was mounted on a UAV, as depicted in the Fig.~\ref{fig:drone_base_fig}a. It served as a transmitter, and the second module was placed at a fixed location, as depicted in the Fig.~\ref{fig:drone_base_fig}b, serving as a receiver for the GCS. 
The communication between the modules was set up over the ESP-NOW communication protocol transmitting data packets on Wi-Fi channel 11, spanning across 22 MHz, with the central frequency $F_0$ = 2.462 GHz. To capture CSI data, the official GitHub repository published by Espressif Systems, containing libraries and binaries to record CSI data on ESP32 boards was used~\cite{esp-csi_21}. To monitor CSI parameters during the communication between a UAV and a GCS, two communication scenarios were established. The first communication scenario, represents communication under static conditions, with both modules remaining immobile at fixed locations at all times. The second communication scenario, represents communication under dynamic conditions, where the transmitter module was mounted onboard a UAV, programmed to fly an automatic mission, while the receiver module remained static at a fixed location. 
\subsection{Jamming Hardware and Software Setup}\label{III-B}
The jammer hardware utilized in the experiments consisted of a Linux-based Dell Latitude E5470 laptop running Ubuntu 20.04 and a programmable SDR from Ettus Research, specifically the USRP B210 with a wideband omnidirectional antenna. 
The USRP B210 provides coverage for frequencies between 70 MHz and 6 GHz, with a max power output of up to 10-17 dBm and 56 MHz of instantaneous bandwidth. This wide frequency coverage makes it suitable for controlled jamming tests in the ISM bands, such as the 2.4 GHz Wi-Fi band. As the primary software framework to design and implement jamming signals, GNU Radio Companion (GRC) was used. GRC is an open-source, Linux-based signal processing runtime and software development toolkit. A custom flow graph was modeled and developed in GRC to emulate a constant jammer, emitting a complex white Gaussian noise sequence with the variable programmable amplitude, sample rate, bandwidth and transmission power values. To achieve the highest jamming effect, upon activation, the jammer was configured to constantly transmit the jamming sequence at the central frequency of the communication channel. By adjusting parameters such as sampling rate, transmission power, covered bandwidth, as well as varying the distance between the jammer and the ESP32-S3 receiver module, different jamming conditions were simulated.

\subsection{Jamming Experiments}\label{III-C}

\paragraph{Static Communication Experiments} Before each experiment, to ensure that there was no intentional or unintentional interference, a spectrogram of the communication channel in the testing environment was recorded, using a Rohde\&Schwarz FSH8 Spectrum Analyzer. After ensuring the communication channel was interference free, the experiments were conducted. At the beginning of each experiment, the transmitter and receier modules were placed at fixed locations. A stable communication link between the transmitter module, mounted on a UAV, and the receiver module at the GCS was established. The transmitter module was constanly sending packets to the rceiver module with the frequency of 100 Hz. To maintain consistency across all experiments, the ESP32-S3 modules were programmed to transmit at a power level of 10 dBm. Afterwards, a constant AWGN jammer was configured according to the communication channel properties. The distance between the jammer and the ESP32-S3 receiver module varied from experiment to experiment, but always remained within 30 meters. The transmission power of the jammer also varied depending on the transmission gain value, programmed in GRC. Due to the fact that USRP B210 SDRs are uncalibrated, specific power output could not be set in dBm, but rather using the transmission gain value. From the conducted experiments it was established that the maximum allowed transmission gain setting of 89.75 dB, stated in the USRP manual, corresponds to the absolute transmission power ranging between 13-17 dBm, which matched with the results published in other studies~\cite{Ajala_22}. Each experiment was separated into two phases: 1\textsuperscript{st} phase - normal communication, and 2\textsuperscript{nd} phase - communication under jamming. In the first half of each experiment, the communication was conducted under normal conditions. In the second half of each experiment, the jammer was activated. The CSI metrics throughout the entirety of each experiment were continuously recorded for further analysis.

\paragraph{Dynamic Communication Experiments} During the dynamic communication experiments, a UAV with the transmitter module was programmed to fly an automatic mission, while the receiver module was placed at a fixed location at the GCS. Communication link and jammer settings remained consistent with the previous experiments conducted in the static scenario. There were two series of experiments conducted. The first series of experiments was conducted under normal conditions without jamming. The second series of experiments was conducted under constant jamming. The CSI metrics were constantly recorded during both series of experiments, for further comparison and analysis.

\section{Experimental Results}

\subsection{Static Communication Experiments}\label{IV-A}
During the static communication experiments, both ESP32-S3 modules were placed outdoors at fixed locations with a clear line of sight between them. The distance between the transmitter and the receiver was kept constant at 50 meters, and no movement or environmental changes were introduced during the experiments, to maintain stable channel conditions. Two static experiments were conducted to test the influence of a jammer on the recorded CSI values. 

In the first experiment, the jammer was placed 10 m away from the ESP32-S3 receiver module and in the second experiment, 30 m away. At the receiver module, CSI data was constantly collected from 52 individual subcarriers of the Wi-Fi communication channel. It included amplitude fluctuation, phase shift, and jitter. In the 1\textsuperscript{st} phase of the first experiment, jammer was inactive, the PDR was recorded at 98.87\%, in the 2\textsuperscript{nd} phase, upon jammer activation, it instantly and completely overwhelmed the receiver, causing a complete loss of communication due to a much higher transmission power.

\begin{figure}[htbp]
\centerline{\centerline{\includegraphics[scale=0.18]{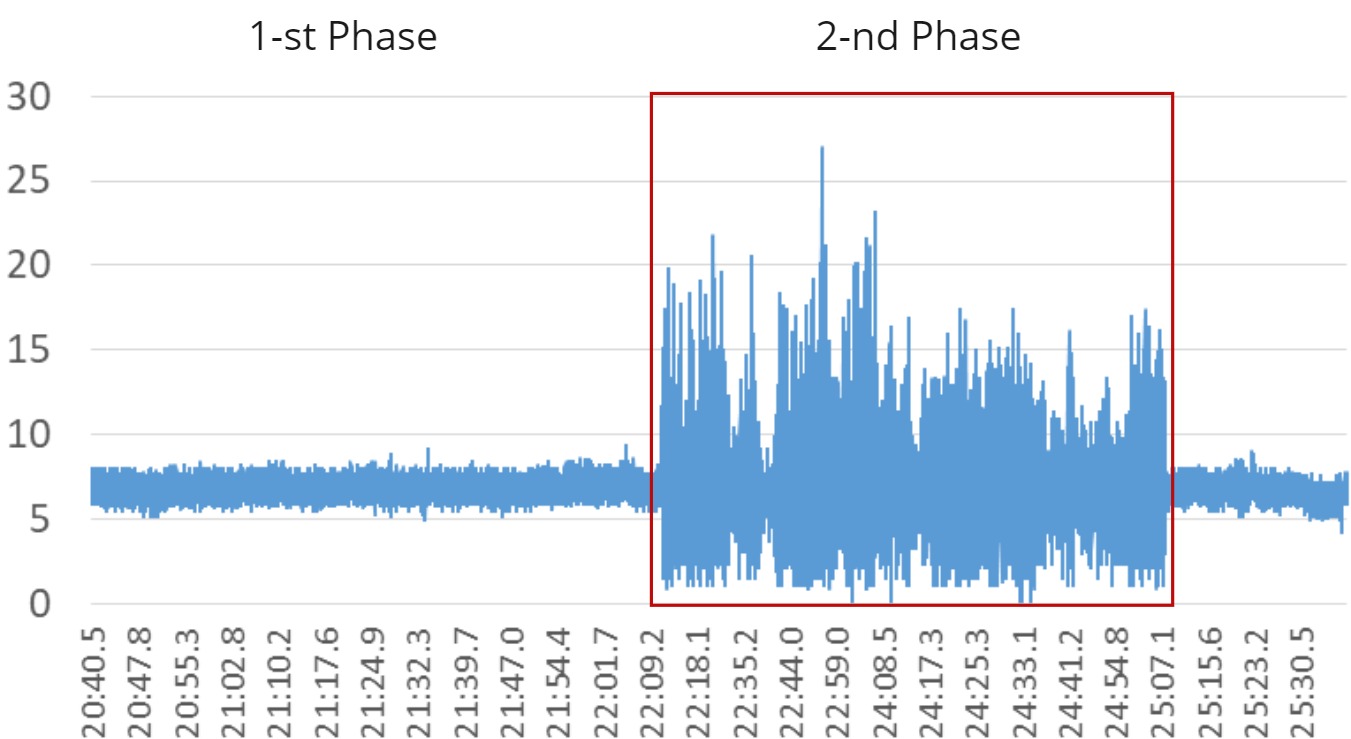}}}
\caption{Plot of the amplitude fluctuation of subcarrier 50 during the second static communication experiment.}
\label{fig:ampl_fig2}
\end{figure}

In the second experiment, the effects of the jammer were less drastic, caused by the attenuation of the jamming signal due to the larger distance to the ESP32-S3 receiver. Fig.~\ref{fig:ampl_fig2} depicts the amplitude fluctuation plot of the signal from subcarrier 50, recorded during the second experiment. The area marked with a red rectangle indicates the part of the experiment where the jammer was active. The active period of the jammer can be clearly identified based on the increase in the amplitude of the signal, compared to the 1\textsuperscript{st} phase, where the jammer was inactive. 

\begin{figure}[htbp]
\centerline{\centerline{\includegraphics[scale=0.18]{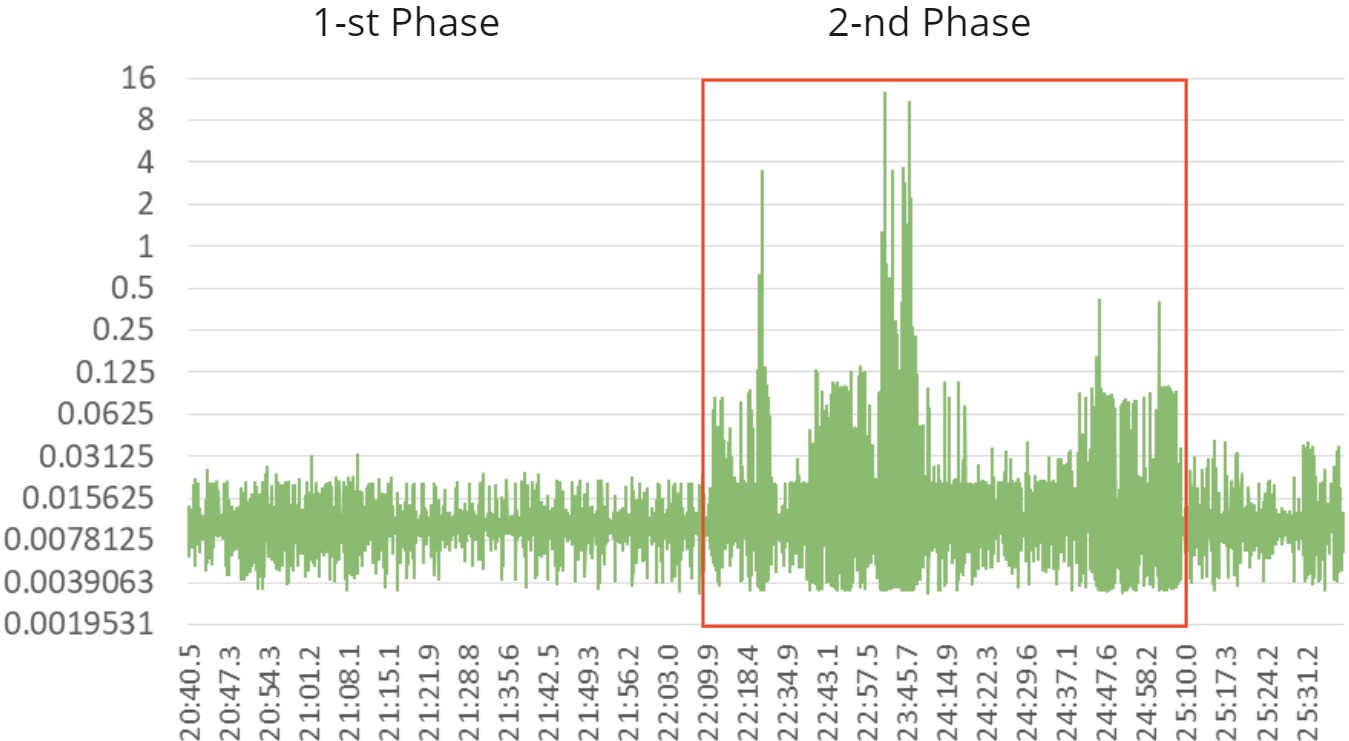}}}
\caption{Plot of the time delay fluctuation of consecutive packets during the second static communication experiment.}
\label{fig:jitter_fig3}
\end{figure}

The activation of the jammer resulted in considerable redistribution of transmission power on multiple subcarriers, but could be observed most prominently on subcarriers 45-52. Additionally, apart from the increased amplitude fluctuation, the increase in time delay on each delivered packet was also clearly observable after the activation of the jammer, as depicted in the Fig.~\ref{fig:jitter_fig3}. The jitter between consecutive packets in the first half of the experiment remained within 5-20 milliseconds, appropriate for the sent packet frequency of 100 Hz. However, upon jammer activation, the jitter variation increased significantly, with delays ranging from 100 milliseconds to several seconds. In some cases, the time delay between consecutive packets was as high as 12.8 seconds, indicating serious disruptions in packet delivery as depicted in the Fig.~\ref{fig:jitter_fig3}. 

 The PDR also dropped drastically after the activation of the jammer. In the 1\textsuperscript{st} phase of the experiment, where the jammer was inactive, PDR was calculated at 97.68\%. In the 2\textsuperscript{nd} phase of the experiment, after jammer activation, PDR decreased to 51.46\%, demonstrating a significant loss of packets. Table~\ref{tab1:experiments} summarises the results from the static communication experiments.
 
\begin{table}[htbp]
\caption{Summary of the results from the static experiments}
\begin{center}
\begin{tabular}{|c|c|c|c|c|}
\hline
\textbf{Condition} & \textbf{Jammer} & \multicolumn{2}{|c|}{\textbf{Channel State Information}} & \textbf{PDR} \\
\cline{3-4}
 & & \textbf{\textit{Amplitude}} & \textbf{\textit{Jitter}} & \\
\hline
\multirow{4}{*}{Static} & Inactive & Low fluct. & 5--20 ms & 98.87\% \\
\cline{2-5}
 & Inactive & Low fluct. & 5--20 ms & 97.68\% \\
\cline{2-5}
 & Active (10 m) & n/a & n/a & n/a \\
\cline{2-5}
 & Active (30 m) & High fluct. & 100 ms--12.8 s & 51.46\% \\
\hline
\end{tabular}
\label{tab1:experiments}
\end{center}
\end{table}

\subsection{Dynamic Communication Experiments}\label{IV-B}
During dynamic communication experiments, the UAV was programmed to fly an automatic mission to eliminate the inconsistencies associated with a manual UAV operation by a remote pilot. Hence, the flight pattern and UAV behavior remained consistent between experiments. The mission was designed to cover a rectangular area, fragmented into 34 individual waypoints. At each waypoint the UAV stopped for~1~second before continuing to the next waypoint, simulating the flight path and behavior of a drone scanning and mapping an agricultural field. The first series of 3 experiments were conducted under normal conditions without jamming. The second series of 3 experiments were conducted under active jamming. The corresponding CSI data was obtained with every experiment.

During the first series of experiments under normal conditions, the amplitude fluctuations of individual subcarriers were noticeably higher, including increased jitter and lower PDR caused by the movement of the UAV, and change of the antenna orientation. Compared to the static experiments, jitter increased by more than 50\%, fluctuating within 10-40 milliseconds, with occasional spikes up to 70 milliseconds, while the lowest recorded PDR was at 84.83\%. This highlights the sensitivity of the communication link to movement and changing spatial antenna configuration. 

The second series of 3 experiment were conducted under active jamming. The transmitted jamming signal maintained the same characteristics as in the previous jamming experiments, and the distance between the jammer and the ESP32-S3 receiver varied between 20 and 30 meters depending on the experiment. Compared to the first series of experiments under normal conditions, there was a change in the amplitude fluctuation range of all of the subcarriers. Overwhelming noise introduced by the jammer affected the CFR, and consequently the recorded CSI values, leading to suppressed amplitude values. Additionally, lost and discarded packets also contributed to the observed effect. The jitter also increased significantly, similar to the static experiments, reaching maximum delays of up to 13.2 seconds between consecutive packets. The PDR also exhibited a significant drop, with lowest recorded PDR of 32.24\%. Table~\ref{tab2:experiments} summarizes the results of the dynamic communication experiments. 

\begin{table}[htbp]
\caption{Summary of the results from the dynamic experiments}
\begin{center}
\begin{tabular}{|c|c|c|c|c|}
\hline
\textbf{Condition} & \textbf{Jammer} & \multicolumn{2}{|c|}{\textbf{Channel State Information}} & \textbf{PDR} \\
\cline{3-4}
 & & \textbf{\textit{Amplitude}} & \textbf{\textit{Jitter}} & \\
\hline
\multirow{6}{*}{Dynamic} & Inactive & High fluct. & 10--70 ms & 90.61\% \\
\cline{2-5}
 & Inactive & High fluct. & 10--100 ms & 88.38\% \\
\cline{2-5}
 & Inactive & High fluct. & 10--120 ms & 84.83\% \\
\cline{2-5}
 & Active (30 m) & Suppressed & 100 ms--7.5 s & 51.84\% \\
\cline{2-5}
 & Active (25 m) & Suppressed & 100 ms--8.6 s & 49.92\% \\
\cline{2-5}
 & Active (20 m) & Suppressed & 100 ms--13.2 s & 32.24\% \\
\hline
\end{tabular}
\label{tab2:experiments}
\end{center}
\end{table}

\section{Conclusion and Future Work}
Across all experiments, static communication conditions were easier to analyze, enabling straightforward detection of changes in CSI metrics. In contrast, dynamic communication conditions introduced volatility in amplitude, phase shift, and time variation jitter, due to the changing antenna angle, UAV movement, and distance increase to the receiver module, requiring more sophisticated analysis to account for it. Static communication experiments under jamming demonstrated that detecting a constant jammer by utilizing such CSI detection parameters as analysis of individual subcarrier amplitude changes, time variation jitter changes and PDR together, is a feasible and reasonable approach. Dynamic communication experiments proved the detection to be more challenging, and required a more thorough analysis through examining standard deviation and variance of the amplitude from individual subcarriers, due to the increased volatility of the communication link. However, even during dynamic communication experiments, the effects of an active constant jammer were observable in the suppressed amplitude values of individual subcarriers, increased jitter, and low PDR. Conducted experiments also showed that CSI analysis allowed for a precise identification of the channel frequencies affected by the jammer, thanks to the fine grained fragmentation of the communication channel into the individual subcarriers. For future work, it is planned to assess the feasibility of CSI analysis to detect other types of jammers, such as random and periodic jammers, and assess its effectiveness.

\section*{Acknowledgment}

This research work was partially funded by the German Federal Office for Information Security through EMiL project under Grant 01MO23014C and the Federal Ministry of Education and Research of Germany through iCampus Upwards II project under Grant  16ES1131.

\bibliographystyle{IEEEtran}

\end{document}